\documentclass[prl,showpacs,twocolumn,amsmath,amssymb]{revtex4}
\usepackage[dvips]{graphicx}
\usepackage{epsfig}
\usepackage{psfrag}

\begin{document}
\title{Spin-zero anomaly in the magnetic quantum oscillations
of a two-dimensional metal}
\author{J. Wosnitza$^1$\cite{presadr}, V. M. Gvozdikov$^2$, J. Hagel$^1$,
P. J. Meeson$^3$\cite{newaddr}, J.~A.~Schlueter$^4$, R. W. Winter$^5$,
and G. L. Gard$^5$}
\affiliation{
$^1$Institut f\"ur Festk\"orperphysik, Technische Universit\"at
Dresden, D-01062 Dresden, Germany\\
$^2$ Max-Planck-Institut f\"ur Physik komplexer Systeme,
D-01187 Dresden, Germany\\
$^3$H. H. Wills Physics Laboratory, University of Bristol, Bristol
BS8 1TL, United Kingdom\\
$^4$Materials Science Division, Argonne National Laboratory, Argonne,
Illinois 60439, USA\\
$^5$Department of Chemistry, Portland State University, Portland,
Oregon 97207}

\date{\today}

\begin{abstract}
We report on an anomalous behavior of the spin-splitting zeros in the
de Haas--van Alphen (dHvA) signal of a quasi-two-dimensional organic
superconductor. The zeros as well as the angular dependence of the
amplitude of the second harmonic deviate remarkably from the standard
Lifshitz--Kosevich (LK) prediction. In contrast, the angular dependence
of the fundamental dHvA amplitude as well as the spin-splitting zeros of
the Shubnikov--de Haas signal follow the LK theory. We can explain this
behavior by small chemical-potential oscillations and find a very good
agreement between theory and experiment. A detailed wave-shape analysis
of the dHvA signal corroborates the existence of an oscillating chemical
potential.
\end{abstract}


\maketitle
For three-dimensional (3D) metals the well-established theory of
Lifshitz and Kosevich (LK) \cite{lif56} can comfortably be utilized
to obtain highly valuable band-structure parameters \cite{sho84}.
The LK theory is well proven and has the advantage of easy applicability
to the experimentally measured magnetic quantum oscillations. The
situation is considerably less resolved for two-dimensional metals.
Both analytical \cite{vag83} as well as numerical \cite{har96} models
have been developed which were proven valid somewhat later by
experiments (see \cite{wos00,har98} and references therein). However,
in these models not all aspects have been taken into account and
they are not as easy applicable as the LK theory. In addition, not
all band-structure parameters can be extracted satisfactorily from
the existing theories leaving some experimental features unexplained.

Prototypical examples for which the fundamental theoretical
predictions can be tested are the quasi-two-dimensional (2D)
organic metals based, e.g., on the organic donor BEDT-TTF (=
bisethylenedithio-tetra\-thia\-fulvalene or ET for short). The
dHvA signal in these layered metals is usually easy to detect
and it is mostly comprised by
only a small number of oscillation frequencies \cite{wos96}.
Consequently, the Fermi surfaces are relatively simple
and in most cases highly two dimensional, i.e., with negligible
dispersion perpendicular to the conducting planes. Nevertheless,
in dHvA signals only seldom notable deviations from the 3D
LK theory appeared \cite{rem1}.

This is remarkably different for the organic superconductor
$\beta{''}$-(BEDT-TTF)$_2$SF$_5$CH$_2$CF$_2$SO$_3$ which shows a
dHvA signal almost perfectly in line with that expected for an
ideal 2D metal with fixed chemical potential \cite{wos00}. However,
some questions remain. That is, in order to fix the chemical potential
either an usually large additional electronic density of states (DOS),
originating from a different band, has to be assumed \cite{wos00} or
some localized states were proposed to be responsible
\cite{nam01}. In addition, although the dHvA signal could be
described extraordinarily well by theory \cite{wos00}
small deviations still are visible (see Fig.\ \ref{2dtheo} below)
\cite{hag01}. Here, we prove this latter feature to be valid
by careful additional measurements utilizing the modulation-field
technique. We further report on an unusual angular dependence of the
spin-splitting zeros of the second harmonic. As we will show explicitly,
both effects reflect the existence of small oscillations of the chemical
potential. Especially the spin-zero anomaly of the second harmonic,
therefore, offers a definite way to validate these oscillations.

We discuss here results of dHvA experiments that have been described
in detail previously \cite{wos00,wos03}. Different high-quality
$\beta{''}$-(BEDT-TTF)$_2$SF$_5$CH$_2$CF$_2$SO$_3$ single crystals
have been measured by use of a capacitance cantilever torquemeter down
to about 0.4~K as well as utilizing the modulation-field technique down
to $\sim$30~mK. The crystals were grown by electrocrystallization at
Argonne National Laboratory \cite{gei96}.

\begin{figure}
\centering
\includegraphics[width=6.5cm,clip=true]{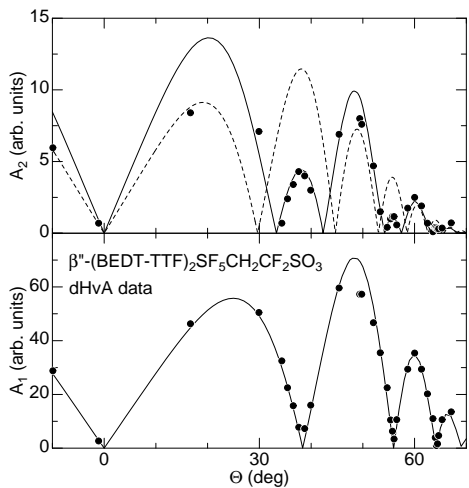}
\caption[]{Angular dependence of the fundamental ($A_1$) and of the
second harmonic ($A_2$) of the dHvA signal of
$\beta{''}$-(BEDT-TTF)$_2$SF$_5$CH$_2$CF$_2$SO$_3$. The solid lines are
obtained by use of (\ref{a1}) and (\ref{a2}) assuming an oscillating chemical
potential. For $A_1$ the same result as in the LK theory is obtained. The
dashed line is the behavior of $A_2$ expected from the LK theory.}
\label{fhdhva}
\end{figure}

For $\beta{''}$-(BEDT-TTF)$_2$SF$_5$CH$_2$CF$_2$SO$_3$, the dHvA signal
consists of only one frequency $F = F_0/\cos(\Theta)$, where $F_0 = (198
\pm 1)$~T is the dHvA frequency at $\Theta = 0$, i.e., for magnetic field
applied perpendicular to the highly conducting plane \cite{wos00,wos03}.
One of the puzzling results we discuss here, is the unusual angular
dependence of the second harmonic, $A_2$, of the dHvA signal
(Fig.\ \ref{fhdhva}) that does not follow the behavior predicted by the
LK theory (dashed line in Fig.\ \ref{fhdhva}). On the other hand,
the fundamental amplitude, $A_1$, is completely in line with expectation.
To be more precise, the dHvA amplitudes in the 2D LK theory are given by
\begin{equation}
A_p = M^0p^{-1}R_T(p)R_D(p)R_S(p),
\end{equation}
where the prefactor $M^0 = \frac{eA}{2\pi^2\hbar}\frac{S(\varepsilon_F)}
{m^\ast}$ is given by the Fermi-surface area $S(\varepsilon_F)$ and the
effective cyclotron mass $m^\ast$, $e$ is the electron charge, $A$ the
sample area, $p$ counts the harmonics, and $R_T(p)$, $R_D(p)$, $R_S(p)$
are the usual damping factors \cite{sho84}. The term dominating the
angular dependence is the spin-splitting factor given by
$R_S(p) = \cos[\frac{1}{2}p\pi g(m^\ast/m_e)]$, where $g$ is the
electron $g$ factor and $m_e$ is the free-electron mass. Since for
the present superconductor $m^\ast/m_e = (2.0\pm 0.1)/\cos(\Theta)$,
$R_S(p)$ repeatedly becomes zero for those angles where the dHvA
oscillations of the spin-up and spin-down electrons interfere destructively.
This allows to determine $gm^\ast/m_e = (3.92\pm 0.01)/\cos(\Theta)$ from
the vanishing of $A_1$ quite accurately. The complete angular dependence
of $A_1$ of the torque signal \cite{rem2} can be well described with a
Dingle temperature $T_D = 0.85$~K for the present sample at temperature
$T = 0.4$~K and magnetic field $B = 14.7$~T (solid line in the lower panel
of Fig.\ \ref{fhdhva}).

In spite of this successful application of the 2D LK theory for $A_1$
it fails clearly to describe the angular dependence of $A_2$ (dashed
line in Fig.\ \ref{fhdhva}). Especially the spin-splitting zeros of
$A_2$ are expected at considerably different positions. Obviously,
some of the assumptions used in the derivation of the 2D LK theory are
not justified for the 2D metal investigated here. Indeed, it has been
predicted that for 2D metals the spin factor $R_S(2)$ may deviate
strongly from the LK behavior \cite{nak00}. Depending on the background
DOS, i.e., the amplitude of chemical-potential oscillation, $A_2$
should vanish at shifted angular positions. The predicted shift is,
however, opposite to what we observe experimentally. That is,
according to \cite{nak00} the first zero of $A_2$ should occur at
smaller angle than given by the LK theory (dashed line in Fig.\
\ref{fhdhva}), the second zero at higher angle and so forth.

It is therefore worthwhile to look for another theoretical explanation.
As was shown previously, the oscillations of the magnetization
in 2D layered conductors and the oscillations of the chemical potential
are closely related and mathematically described by similar (but not
identical) series \cite{gvo03}. The oscillating part of the magnetization
can be written as \cite{remeq2}
\begin{equation}
\tilde{M} = M^0{\rm{Im}}\sum\limits_{p=1}^{\infty}
\frac{(-1)^{p}}{p}\exp \left[ 2\pi i p\left( \frac{F}{B}+
\frac{\tilde{\mu}}{\hbar \omega_c}\right) \right]\hat{R}(p),
\label{mosc}
\end{equation}
with $F=S(\varepsilon_F)/(2\pi e\hbar)$, the cyclotron frequency
$\omega_c = eB/m^\ast$, and $\hat{R}(p)=I(p)R_T(p)R_S(p)R_D(p)$.
The factor $I(p)$ takes account of interlayer electron-hopping
effects which are beyond the LK theory \cite{gvo84}. The oscillating
part of the chemical potential, $\tilde{\mu}=\mu-\varepsilon_F$, is
given by
\begin{equation}
\tilde{\mu} = \hbar \omega_c{\rm{Im}}\sum\limits_{p=1}^{\infty }
\frac{(-1)^{p}}{\pi p}\exp \left[ 2\pi i p\left( \frac{F}{B}+
\frac{\tilde{\mu}}{\hbar \omega_c}\right) \right]\hat{R}(p).
\label{muosc}
\end{equation}
The consequences of $\tilde{\mu} \neq 0$ can be realized by
considering the oscillating correction to $\mu$ in lowest order
\begin{equation}
\mu=\varepsilon_F-\eta\sin \left(2\pi\frac{F}{B}\right).
\label{mu1st}
\end{equation}
We assume here that $\eta=\hbar\omega_c\hat{R}(1)/\pi \ll 1$ is a
small real parameter. Using the notations $z_p=2\pi p F/B$,
$\eta_p=2\pi p \eta/\hbar\omega_c$, and the identity
\begin{equation}
\exp(-i\eta_p\sin z_1)=\sum_{n=-\infty}^{\infty}(-1)^n J_n(\eta_p)
\exp(iz_n),
\label{identity}
\end{equation}
where $J_n(\eta_p)$ is the Bessel function of order $n$,
one can write the magnetization (\ref{mosc}) in the standard form
\begin{equation}
\tilde{M}=\sum_{n=1}^{\infty}A_n\sin\left(2\pi n \frac{F}{B}\right).
\label{mstandard}
\end{equation}
The amplitudes of the harmonics are given by
\begin{equation}
A_n=M^0\sum\limits_{p=1}^{\infty}\frac{(-1)^{p}}{p}\hat{R}(p)
\left[J_{p+n}(\eta_p)-J_{p-n}(\eta_p)\right].
\label{harmonics}
\end{equation}
The amplitudes $A_n$ are, therefore, weighted sums of the terms
$\hat{R}(p)$. Contrary to the LK theory the latter term contains
the additional factor $I(p)$ that takes into account such effects
as interlayer hopping \cite{gvo03,gvo84} or the dispersion of
magnetic-breakdown bands \cite{gvo02}. (For
$\beta{''}$-(BEDT-TTF)$_2$SF$_5$CH$_2$CF$_2$SO$_3$ magnetic breakdown
is irrelevant as shown by measurements up to 60~T \cite{wos01}.)
Important consequences of (\ref{harmonics}) are deviations from
the usual LK temperature and magnetic-field dependences. This is
realized, e.g., in the effective masses which apparently become
smaller for each higher harmonic when extracted by use of the LK
formula \cite{sho84}. In the present case, an apparent effective mass
of only about 1.5~$m_e$ is obtained for the second harmonic.

What is of importance here, is that the angular dependence, i.e., the
spin-zero positions also differ from those predicted in the LK approach.
To show this in more detail we consider the fundamental and second
harmonic which can be compared to the experimental data. Since we
assume $\eta \ll 1$, also $\eta_1$ and $\eta_2$ are small parameters.
Accordingly, the relevant Bessel functions in (\ref{harmonics}) can
be approximated as $J_0(\eta_1)\approx J_0(\eta_2)\approx 1$ and
$J_1(\eta_1)\approx \eta_{1}/2=\hat{R}(1)$ resulting in
\begin{eqnarray}
A_1 &= &M^0\hat{R}(1),\label{a1}\\
A_2 &= &-M^0[\frac{1}{2}\hat{R}(2) + \hat{R}(1)^2].
\label{a2}
\end{eqnarray}
\begin{figure}
\centering
\includegraphics[width=8.3cm,clip=true]{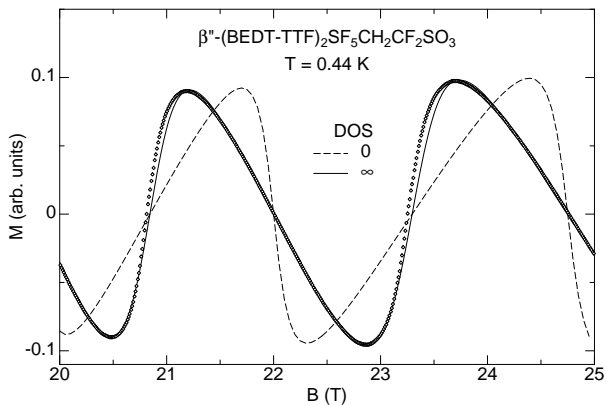}
\caption[]{Comparison of the measured dHvA data (open symbols) with
the calculated signals for a 2D metal with fixed number of charge
carriers (dashed line) and for a 2D metal with fixed chemical
potential (solid line).}
\label{2dtheo}
\end{figure}
For a weakly oscillating chemical potential, therefore, the amplitude
of the fundamental basically remains identical to the LK prediction,
whereas $A_2$ becomes a linear combination of the damping factors for
$p = 1$ and $p = 2$. Consequently, care has to be taken when extracting
band-structure parameters from the second harmonic. Besides the
modified temperature dependence (leading to the above mentioned
apparent effective-mass peculiarities), an unusual angular dependence
of $A_2$ with shifted spin-splitting zeros results. The latter are
determined by $[\frac{1}{2}\hat{R}(2)+\hat{R}(1)^2]=0$. With $R_D(1)^2=
R_D(2)$ and $R_S(p)$ as stated above, the spin-splitting zeros are
given by
\begin{equation}
\cos\left(\frac{\pi g m^\ast}{m_e}\right)I(2)+
2\cos^2 \left(\frac{\pi g m^\ast}{2m_e}\right)
\frac{R^2_T(1)}{R_T(2)} I^2(1)=0.
\label{ssz}
\end{equation}
Thus, the zeros are shifted as compared to the LK theory where the second
term is absent. The shift is a weak function of temperature and magnetic
field caused by the factors $R_T(p)$ and $I(p)=\int g(\varepsilon)\exp(2\pi
ip\varepsilon/\hbar \omega_c) {\rm d}\varepsilon$. The layer-stacking
factor for a simple cosine-like interlayer dispersion can be written as
$I(p)=J_0(\frac{4\pi tp}{\hbar \omega_c})$, where $t$ is the
interlayer-hopping integral. For the present 2D superconductor
there is no detectable dispersion across the layers and the DOS
$g(\varepsilon)$ associated with the electron hopping between
the layers is unknown \cite{wos02}. Nonetheless, because the hopping
integral is very small compared to $\hbar\omega_c$, $I(p)$ may
be approximated by 1.

The excellent agreement between our theory and experiment is evident
from Fig.\ \ref{fhdhva} where we used Eqs.\ (\ref{a1}) and (\ref{a2})
to obtain the solid lines. Since all experimental parameters, $m^\ast$,
$T$, $B$, and $T_D$, are well known there is no free parameter except
for simple scaling factors.

\begin{figure}
\centering
\includegraphics[width=8cm,clip=true]{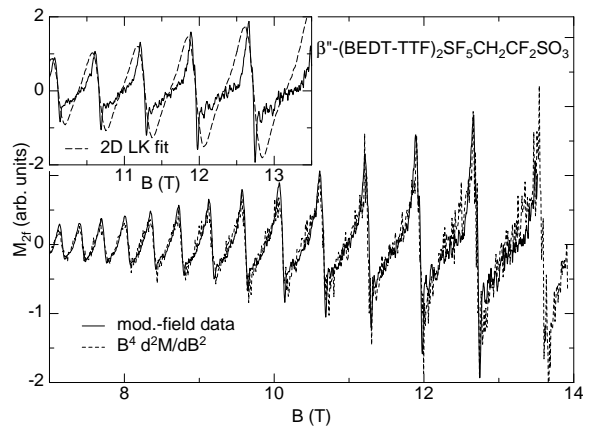}
\caption[]{Comparison of modulation-field dHvA data with the second
derivative of magnetization data times $B^4$ measured by the torque
method. The inset shows the modulation-field data in comparison to
the expected signal for a 2D metal with fixed chemical potential.}
\label{mvsx}
\end{figure}

This result implies that a weak oscillation of the
chemical potential exists. Indeed, when analyzing in detail the
experimental dHvA wave shape small deviations from the 2D LK behavior
for fixed chemical potential can be resolved. In Fig.\ \ref{2dtheo}
it is obvious that the observed steep increase of the
dHvA signal cannot be described satisfactorily by the theoretical 2D
LK behavior (solid line in Fig.\ \ref{2dtheo}). This corroborates
the notion of an oscillating chemical potential. This oscillation,
however, must be different from the usually predicted
sawtooth-like 2D behavior \cite{vag83} as visualized by the dashed
line in Fig.\ \ref{2dtheo}.

Since possible artifacts, such as torque interaction, might obscure
the dHvA signal we checked the validity of our torque result by
comparing it with modulation-field data (Fig.\ \ref{mvsx}).
For a modulation-field amplitude not too large and signal detection
on the second harmonic, the modulation-field data are approximately
proportional to the second derivative of the magnetization with
respect to $B$ times $B^4$ \cite{sho84}. The excellent
agreement between both signals is evident \cite{rem3}.
This proves the validity of both experimental data and verifies
the deviation from the 2D LK behavior as real. Indeed, for the
modulation-field data the deviation appears even more pronounced
since the second derivative of $M$ is analyzed (inset of
Fig.\ \ref{mvsx}).

Consequently, these results substantiate the existence of an oscillating
chemical potential. This oscillation is small, as assumed and realized
by the minute wave-shape effects, but must be more elaborate
than the simple lowest-order sinusoidal waveform considered in
(\ref{mu1st}). Qualitatively, the waveform of $\tilde{\mu}$ must be
fast changing for rising $M$ and slowly varying for decreasing $M$,
similar as visualized in Fig.\ 1 of Ref.\ \cite{har96} for an almost
fixed chemical potential.
Indeed, the actual field dependence of the chemical-potential
oscillations might be extractable by an iterative
fitting procedure using Eqs.\ (\ref{mosc}) and (\ref{muosc}).

\begin{figure}
\centering
\includegraphics[width=6.5cm,clip=true]{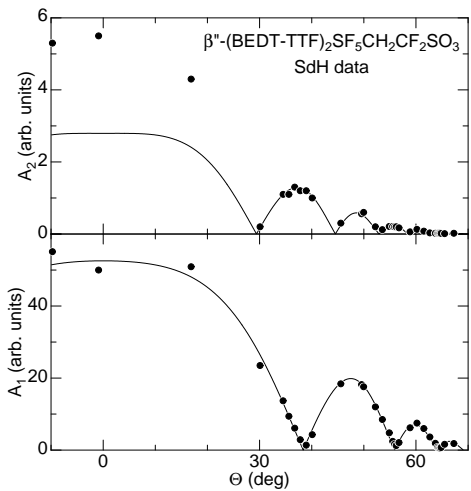}
\caption[]{Angular dependence of the fundamental ($A_1$) and the second
harmonic ($A_2$) of the SdH signal. The solid lines show the expected
behavior according to the 2D LK theory.}
\label{fhsdh}
\end{figure}

The final strong argument in favor for an oscillating chemical potential
is the absence of any anomalous shifts of the spin-splitting zeros in
the SdH signal (Fig.\ \ref{fhsdh}). Here, for the same parameters, $T$,
$B$, and $T_D$, as for the dHvA data in Fig.\ \ref{fhdhva}, the 2D LK
theory describes the angular dependences of the SdH amplitudes very
well (except for some $A_2$ points close to $\Theta = 0$).
In particular, the zeros of the second harmonic lie exactly at the
positions expected for a metal with fixed chemical potential.
Obviously, the electrical leads, necessary for measuring the SdH
signal, act as a charge-carrier reservoir leading to a constant
chemical potential. It is noteworthy that for an
(even with small amplitude) oscillating chemical potential of
inverse-sawtooth shape split peaks in the SdH signal should
occur \cite{gvo04}. In line with a fixed chemical potential for
electrical-transport measurements, however, such split peaks do
not occur in our SdH measurements.

In conclusion, we observed and explained quantitatively an anomalous
angular dependence of the dHvA signal in a 2D metal. This is shown
to be a genuine effect of the two dimensionality that can be utilized
as a direct proof for an oscillating chemical potential. In the present
case these oscillations are very small but directly visible in the
detailed dHvA wave shape. In SdH experiments, no chemical-potential
oscillations are detected which in turn means that charge oscillates
into and out of the sample during a SdH period.

Part of this work was supported by INTAS, project INTAS-01-0791,
the NATO Collaborative Linkage Grant No.\ 977292, and by the ESF
Scientific Programme on Fermi-liquid instabilities in correlated
metals (FERLIN).
Work at Argonne National Laboratory was supported
by the U.S.~Dept.\ of Energy (W-31-109-ENG-38).
Work at Portland State University was supported by NSF
(Che-9904316).
V.M.G. is grateful to P.\ Fulde and S.\ Flach for the hospitality
at the MPIPKS in Dresden.


\begin{thebibliography}{123}
\bibitem[*]{presadr}
Present address: Forschungszentrum Rossendorf, Hoch\-feld-Magnetlabor
Dresden (HLD), 01314 Dresden, Germany

\bibitem[\dag]{newaddr}
Present address: Royal Halloway, University of London, Egham,
Surrey TW20 0EX, United Kingdom

\bibitem{lif56}
I. M. Lifshitz and A. M. Kosevich, Zh. Eksp. Teor. Fiz. {\bf 29}, 730
(1955) [Sov. Phys. JETP {\bf 2}, 636 (1956)].

\bibitem{sho84}
D. Shoenberg, {\it Magnetic Oscillations in Metals} (Cambridge University
Press, Cambridge 1984).

\bibitem{vag83}
I.D. Vagner {\it et al.},
Phys. Rev. Lett. {\bf 51},
1700 (1983).

\bibitem{har96}
N. Harrison {\it et al.},
Phys. Rev. B {\bf 54}, 9977 (1996).

\bibitem{wos00}
J. Wosnitza {\it et al.},
Phys. Rev. B {\bf 61}, 7383 (2000).

\bibitem{har98}
N. Harrison {\it et al.},
Phys. Rev. B {\bf 58}, 10\,248 (1998).

\bibitem{wos96}
J. Wosnitza, {\it Fermi surfaces of Low-Dimensional Organic Metals and
Superconductors} (Springer, Berlin, 1996).

\bibitem{rem1}
Strong deviations from 3D LK behavior were reported for the SdH signals
in some organic metals. In magnetotransport data, however, some unique
phenomena may obscure the real thermodynamic behavior. See
\cite{wos00} and J. Wosnitza {\it et al.},
Phys. Rev. Lett. {\bf 86}, 508 (2001).

\bibitem{nam01}
In M.-S. Nam {\it et al.},
Phys. Rev. Lett. {\bf 87}, 117001 (2001) localized states caused by a
density-wave state have been suggested. This has been assumed from
a resistive maximum found during cooling in one sample. In contrast,
in all our samples we observed a metallic resistivity leaving
the existence of a density-wave state highly questionable.

\bibitem{hag01}
J. Hagel {\it et al.},
Synth. Metals {\bf 120}, 813 (2001).

\bibitem{wos03}
J. Wosnitza {\it et al.},
Phys. Rev. B {\bf 67}, 060504(R) (2003).

\bibitem{gei96}
U. Geiser {\it et al.},
J. Am. Chem. Soc. {\bf 118}, 9996 (1996).

\bibitem{rem2}
In torque dHvA measurements the additional factor $\sin(\Theta)/
\cos^2(\Theta)$ occurs for the present 2D Fermi surface leading
to zero amplitudes at $\Theta = 0$.

\bibitem{nak00}
M. Nakano, Phys. Rev. B {\bf 62}, 45 (2000).

\bibitem{gvo03}
V.M. Gvozdikov {\it et al.},
Phys. Rev. B {\bf 68}, 155107 (2003).

\bibitem{remeq2}
For $\tilde{\mu} = 0$ and $I(p) = 1$, Eq.\ (\ref{mosc}) is nothing
but the 2D LK theory. In the present case $I(p) \approx 1$
may be used.

\bibitem{gvo84}
V.M. Gvozdikov, Sov. Phys. Solid State {\bf 26}, 1560 (1984).

\bibitem{gvo02}
V.M. Gvozdikov {\it et al.},
Phys. Rev. B {\bf 65}, 1651 (2002).

\bibitem{wos01}
J. Wosnitza {\it et al.},
Physica B {\bf 294-295}, 442 (2001).

\bibitem{wos02}
J. Wosnitza {\it et al.},
Phys. Rev. B {\bf 65}, 180506(R) (2002).

\bibitem{rem3}
The modulation-field data were taken at 30~mK, the torque
data at 0.44~K. Slightly different Dingle temperatures
($T_D = 0.4$~K of the torque sample and $T_D = 0.58$~K for the
modulation-field sample) partly compensate this and lead to the
particularly good agreement.

\bibitem{gvo04}
V.M. Gvozdikov, Phys. Rev. B {\bf 70}, 085113 (2004).

\end{thebibliography}
\end{document}